\documentclass[a4paper,11pt]{article}
\usepackage[left=20mm,right=20mm,top=20mm,bottom=20mm]{geometry}
\usepackage{amsmath}
\usepackage{amsfonts}
\usepackage{amssymb}
\usepackage{amsbsy}
\usepackage{color,graphics}
\usepackage{setspace}
\usepackage{epsfig}
\usepackage{subfigure}
\usepackage{cite}
\usepackage{graphicx}
\usepackage{latexsym,multicol}
\usepackage{tcolorbox}
\begin{document}
\title{\huge\textsf{\textbf{Deformation localisation in stretched\\ liquid crystal elastomers}}}
\author{Rabin Poudel$^{*}$\qquad
	Yasemin \c{S}eng\"{u}l$^{*}$\qquad
L. Angela Mihai$^{*}$\\ \\
\textit{$^{*}${School of Mathematics, Cardiff University,  Cardiff, UK}	}}
%\date{}
\maketitle

\begin{abstract}\vskip 6pt
\hrule\vskip 12pt
We model within the framework of finite elasticity two inherent instabilities observed in liquid crystal elastomers under uniaxial tension. First is necking which occurs when a material sample suddenly elongates more in a small region where it appears narrower than the rest of the sample. Second is shear striping, which forms when the in-plane director rotates gradually to realign and become parallel with the applied force. These phenomena are due to the liquid crystal molecules rotating freely under mechanical loads. To capture necking, we assume that the uniaxial order parameter increases with tensile stretch, as reported experimentally during polydomain-monodomain transition. To account for shear striping, we maintain the uniaxial order parameter fixed, as suggested by experiments. Our finite element simulations capture well these phenomena. As necking in liquid crystal elastomers has not been satisfactorily modelled before, our theoretical and numerical findings related to this effect can be of wide interest. Shear striping has been well studied, yet our computed examples also show how optimal stripe width increases with the nematic penetration depth measuring the competition between the Frank elasticity of liquid crystals and polymer elasticity.  Although known theoretically, this result has not been confirmed numerically by previous nonlinear elastic models.   \\

 \vskip 12pt
\hrule
\end{abstract}

%\tableofcontents

%%%%%%%%%%%%%%%%%%%%%%%%%%%%%%%%%%%%%%%%%%%%%%%%%%%%%%%%%%%%
%%%%%%%%%%%%%%%%%%%%   NEW SECTION  %%%%%%%%%%%%%%%%%%%%%%%%
%%%%%%%%%%%%%%%%%%%%%%%%%%%%%%%%%%%%%%%%%%%%%%%%%%%%%%%%%%%%
\section{Introduction}

Liquid crystal elastomers (LCEs) are formed from cross-linked polymeric chains with embedded liquid crystal (LC) molecules \cite{Warner:2007:WT}. Due to this special architecture, they display both self-organisation under external stimuli, such as heating or illumination, and large reversible deformations under mechanical loads. In modern laboratories, LCEs are being prepared through various techniques. Typically, polydomain samples are obtained where the material contains multiple subdomains, with a different nematic alignment. The unit vector for the localised direction of uniaxial nematic alignment is termed the \emph{director}. Monodomain LCEs can subsequently be achieved by applying electric or magnetic fields, or deforming the material mechanically to induce a desired nematic orientation. 

There has been a long-standing interest in the multi-physics characteristics of nematic elastomers and incorporating them in muscle-like actuators and other advanced technologies \cite{deGennes:1997:etal,Kim:2023:KLK,Rothemund:etal:2021,White:2015:WB}. However, many fundamental questions concerning these material remain open and require further theoretical investigation.

\begin{figure}[htbp]
	\begin{center}
		\includegraphics[width=0.5\textwidth]{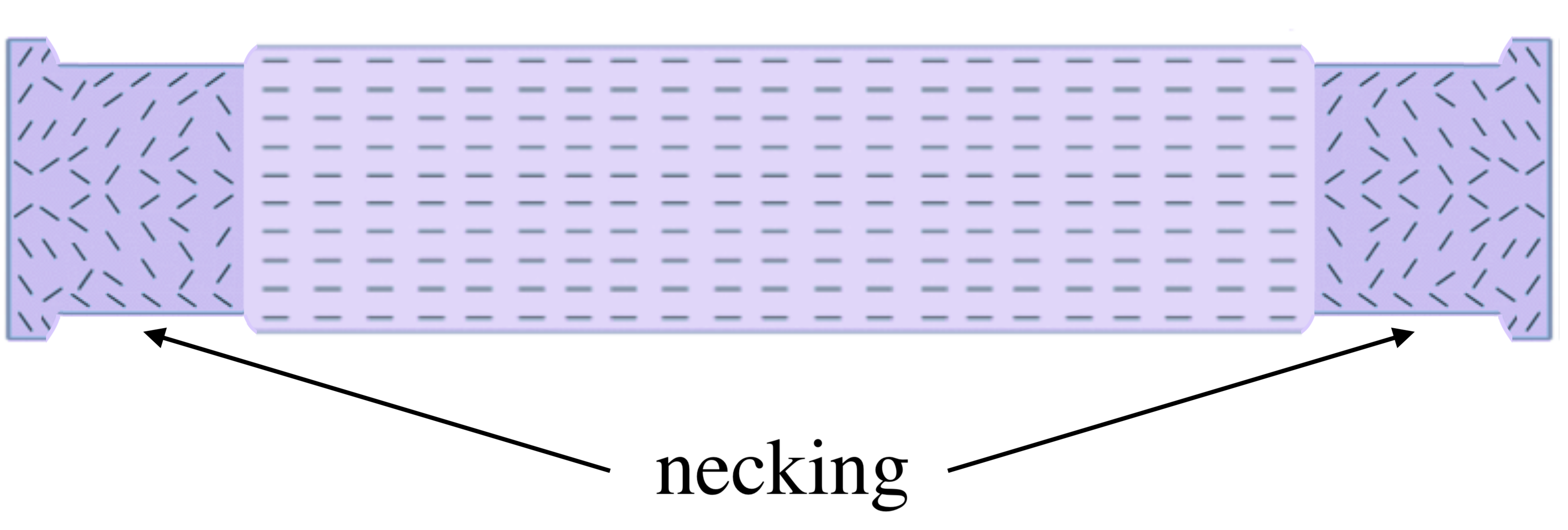}
		\caption{Schematics of nematic LCE developing necking under large stretch causing the initially disordered LC molecules to rotate until they align with the applied force in the horizontal direction. Darker colour at the ends signifies opacity in the physical sample where the LC molecules are randomly oriented, while the lighter colour in the middle corresponds to the physical sample being translucent where the LC molecules are uniformly aligned.}\label{fig:nematic-necking}
	\end{center}
\end{figure}

Generally, under mechanical loads, the nematic director tends to rotate in order to align with the largest principal stretch \cite{Mihai:2022,Mitchell:MDG:1993}. This rotation is not always uniform and can generate some interesting macroscopic effects. For example, when subject to a large tensile force, some LCEs exhibit localised necking \cite{Clarke:1998:CT,Clarke:1998:CTKF}. This phenomenon occurs when there is a critical extension ratio, such that the force required to extend the material beyond this critical value changes from increasing to decreasing. In this case, the homogeneous deformation becomes unstable, and the material sample suddenly elongates more in a small region than in the rest of the sample. Locally, the material appears much narrower than before the critical stretch was reached, as illustrated schematically in Figure~\ref{fig:nematic-necking}. Experimental evidence of LCE necking has been reported in \cite{He:2020:HZHC,Higaki:2013:HTU,Li:2022:etal}. Also, in \cite{He:2020:HZHC}, since necking could not be captured by the neoclassical LCE model based on the neo-Hookean strain-energy function for rubber, it was suggested that a viscoelastic model would be required instead. A finite element modelling of instabilities in viscoelastic nematic elastomers, including necking, is presented in \cite{Chehade:2024:etal}. However, within the theoretical framework of elasticity, a satisfactory mathematical model for LCE necking  is still to be achieved.

%%%%%%%%%%%%%%%%
%%%%%%%%%%%%%%%

\begin{figure}[htbp]
	\begin{center}
		\includegraphics[width=0.9\textwidth]{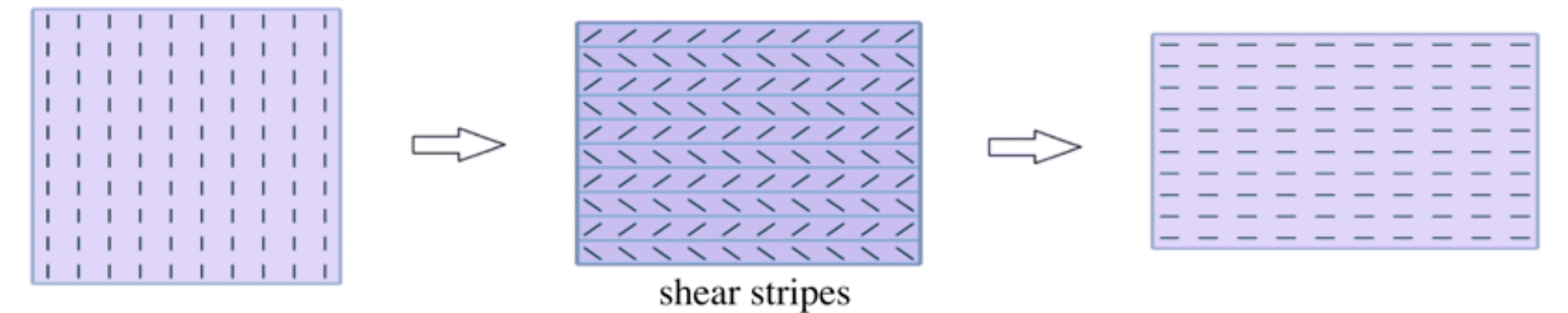}
		\caption{Schematics of nematic LCE developing shear stripes under large stretch causing the initial director in the vertical direction to rotate until it aligns with the applied force in the horizontal direction. Darker colour in the middle figure signifies opacity in the physical sample where the shear stripes form, while lighter colour in the left- and right-hand figures corresponds to the physical sample being translucent where the LC molecules are uniformly aligned.}\label{fig:nematic-stripes}
	\end{center}
\end{figure}

Certain stretch deformations of LCEs also result in a pattern of parallel stripe subdomains where equal and opposite shear deformations occur in neighbouring stripes \cite{Golubovic:1989:GL,Kundler:1995:KF}. The theoretical explanation for this phenomenon, which is depicted schematically in Figure~\ref{fig:nematic-stripes}, is that the energy depending isotropically on the macroscopic deformation through the relative strain of the microstructure is minimised by a state of many homogeneously deformed parts. This is known as \emph{soft} or \emph{semi-soft elasticity}, since it is usually accompanied by a much slower increase in the required stress compared to prior and subsequent deformed states  \cite{Carlson:2002:CFS}. Shear stripes formation in elongated nematic LCEs  \cite{Finkelmann:1997:FKTW,Higaki:2013:HTU,Kundler:1995:KF,Kundler:1998:KF,Petelin:2009:PC,Petelin:2010:PC,Talroze:1999:etal,Zubarev:1999:etal} has been extensively analysed theoretically  \cite{Bladon:1993:BTW,Bladon:1994:BTW,Carlson:2002:CFS,Conti:2002a:CdSD,DeSimone:2000:dSD,DeSimone:2009:dST,Fried:2004:FS,Fried:2005:FS,Fried:2006:FS,Kundler:1995:KF,Mihai:2022,Mihai:2020a:MG,Mihai:2021a:MG,Mihai:2023,Mihai:2023:MRGMG} or simulated numerically \cite{Mbanga:2010:MYSS,Soltani:etal:2021}, and it is therefore well understood.

%%%%%%%%%%%%%%%%
%%%%%%%%%%%%%%%

In this paper, first we introduce a theoretical model for LCEs capable of necking under uniaxial tensile load (Section~\ref{sec:necking}). We then revisit the modelling of LCE shear striping under uniaxial tension, providing an analytical solution and estimating stripe periodicity \cite{Verwey:1996:VWT} (Section~\ref{sec:stripes}). These models are further implemented in a finite element code to simulate the respective effects (Section~\ref{sec:numerics}). Necking in LCEs has not been satisfactorily modelled before, so both our theoretical and numerical findings related to this effect can be of wide interest. While shear striping has been well studied, our numerical examples also show how optimal stripe width increases with the nematic penetration depth measuring the competition between the Frank elasticity of LCs and polymer elasticity. This result, although known theoretically, has not been captured numerically by previous nonlinear elastic models. 

%%%%%%%%%%%%%%%%%%%%%%%%%%%%%%%%%%%%%%%%%%%%%%%%%%%%%%%%%%%%
%%%%%%%%%%%%%%%%%%%%   NEW SECTION  %%%%%%%%%%%%%%%%%%%%%%%%
%%%%%%%%%%%%%%%%%%%%%%%%%%%%%%%%%%%%%%%%%%%%%%%%%%%%%%%%%%%%
\section{Necking}\label{sec:necking}

To describe an incompressible nematic LCE, we consider the following strain-energy function \cite{Fried:2006:FS}
\begin{equation}\label{eq:Wlce:nh:nc}
	W^{(lce)}=\frac{\mu^{(1)}}{2}\left(\lambda_{1}^{2}+\lambda_{2}^{2}+\lambda_{3}^{2}-3\right)
	+\frac{\mu^{(2)}}{2}\left(\alpha_{1}^{2}+\alpha_{2}^{2}+\alpha_{3}^{2}-3\right),
\end{equation}
where $\mu=\mu^{(1)}+\mu^{(2)}>0$ is the shear modulus at infinitesimal strain, with $\mu^{(1)}, \mu^{(2)}\geq0$ constant material parameters, $\{\lambda_{1}^2,\lambda_{2}^2,\lambda_{3}^2\}$ denote the eigenvalues of $\textbf{F}\textbf{F}^{T}$, with $\textbf{F}$ the deformation gradient from the reference cross-linking state, satisfying $\det\textbf{F}=1$, and $\{\alpha_{1}^2,\alpha_{2}^2,\alpha_{3}^2\}$ represent the eigenvalues of $\textbf{A}\textbf{A}^{T}$, with $\textbf{A}=\textbf{G}^{-1}\textbf{F}\textbf{G}_{0}$ the local elastic deformation tensor satisfying $\det\textbf{A}=1$, while $\textbf{G}_{0}$ and $\textbf{G}$ are the natural deformation tensors due to the nematic director in the reference and current configuration, respectively. 

Assuming that the LCE is intrinsically uniaxial, the natural deformation tensor takes the form
\begin{equation}\label{NLC:eq:G}
	\textbf{G}=a^{-1/6}\textbf{I}+\left(a^{1/3}-a^{-1/6}\right)\textbf{n}\otimes\textbf{n},
\end{equation}
where $\textbf{n}$ is the local nematic director, $\otimes$ denotes the tensor product of two vectors, $\textbf{I}=\mathrm{diag}(1,1,1)$ is the identity tensor, and 
\begin{equation}\label{NLC:eq:aQ}
	a=\frac{1+2Q}{1-Q}
\end{equation}
represents the natural shape parameter, with $Q$ the scalar uniaxial order parameter ($Q=1$ corresponds to perfect nematic order and $Q=0$ to the case where mesogens are randomly oriented) \cite{deGennes:1993:dGP}. For the reference configuration, $\textbf{G}$ is replaced by $\textbf{G}_{0}$, with $\textbf{n}_{0}$, $a_{0}$ and $Q_{0}$ instead of $\textbf{n}$, $a$ and $Q$, respectively. 

Note that, when $\mu^{(2)}=0$, the phenomenological model in equation \eqref{eq:Wlce:nh:nc} reduces to the neo-Hookean model for rubber \cite{Treloar:1944}, and when $\mu^{(1)}=0$, it simplifies to the neoclassical model for LCEs \cite{Bladon:1993:BTW,Bladon:1994:BTW}.

The nematic elastomer modelled by equation \eqref{eq:Wlce:nh:nc} is subject to the finite-strain deformation with $\lambda_{1}=\lambda>1$ and $\lambda_{2}=\lambda_{3}=\lambda^{-1/2}$, while the tensile force is parallel to the nematic director in the current configuration, so that
\begin{equation}
	\textbf{G}=\mathrm{diag}\left(\left(\frac{1+2Q}{1-Q}\right)^{1/3},\left(\frac{1-Q}{1+2Q}\right)^{1/6},\left(\frac{1-Q}{1+2Q}\right)^{1/6}\right).
\end{equation}
Taking $\textbf{G}_{0}=\textbf{I}$, we have
\begin{equation}
	\lambda_{1}=\alpha_{1}\left(\frac{1+2Q}{1-Q}\right)^{1/3},\qquad
	\lambda_{2}=\alpha_{2}\left(\frac{1-Q}{1+2Q}\right)^{1/6},\qquad
	\lambda_{3}=\alpha_{3}\left(\frac{1-Q}{1+2Q}\right)^{1/6},
\end{equation}
and
\begin{equation}
	\alpha_{1}=\lambda\left(\frac{1+2Q}{1-Q}\right)^{-1/3}=\alpha,\qquad
	\alpha_{2}=\alpha_{3}=\lambda^{-1/2}\left(\frac{1-Q}{1+2Q}\right)^{-1/6}=\alpha^{-1/2}.
\end{equation}

For uniaxial tension, the principal components of the first Piola-Kirchhoff stress tensor are, respectively,
\begin{eqnarray}
	P_{1}&=&\frac{\partial W^{(lce)}}{\partial\lambda_{1}}-p\lambda_{1}^{-1}=P>0,\\
	P_{2}&=&\frac{\partial W^{(lce)}}{\partial\lambda_{2}}-p\lambda_{2}^{-1}=0,\\ 	
	P_{3}&=&\frac{\partial W^{(lce)}}{\partial\lambda_{3}}-p\lambda_{3}^{-1}=0,
\end{eqnarray}
where  $p$ is the Lagrange multiplier for the incompressibility constraint, also known as the \emph{arbitrary hydrostatic pressure} \cite{Mihai:2022}. 

%%%%%%%%%%%%%%%%
\begin{figure}[htbp]
	\begin{center}
		\includegraphics[width=0.99\textwidth]{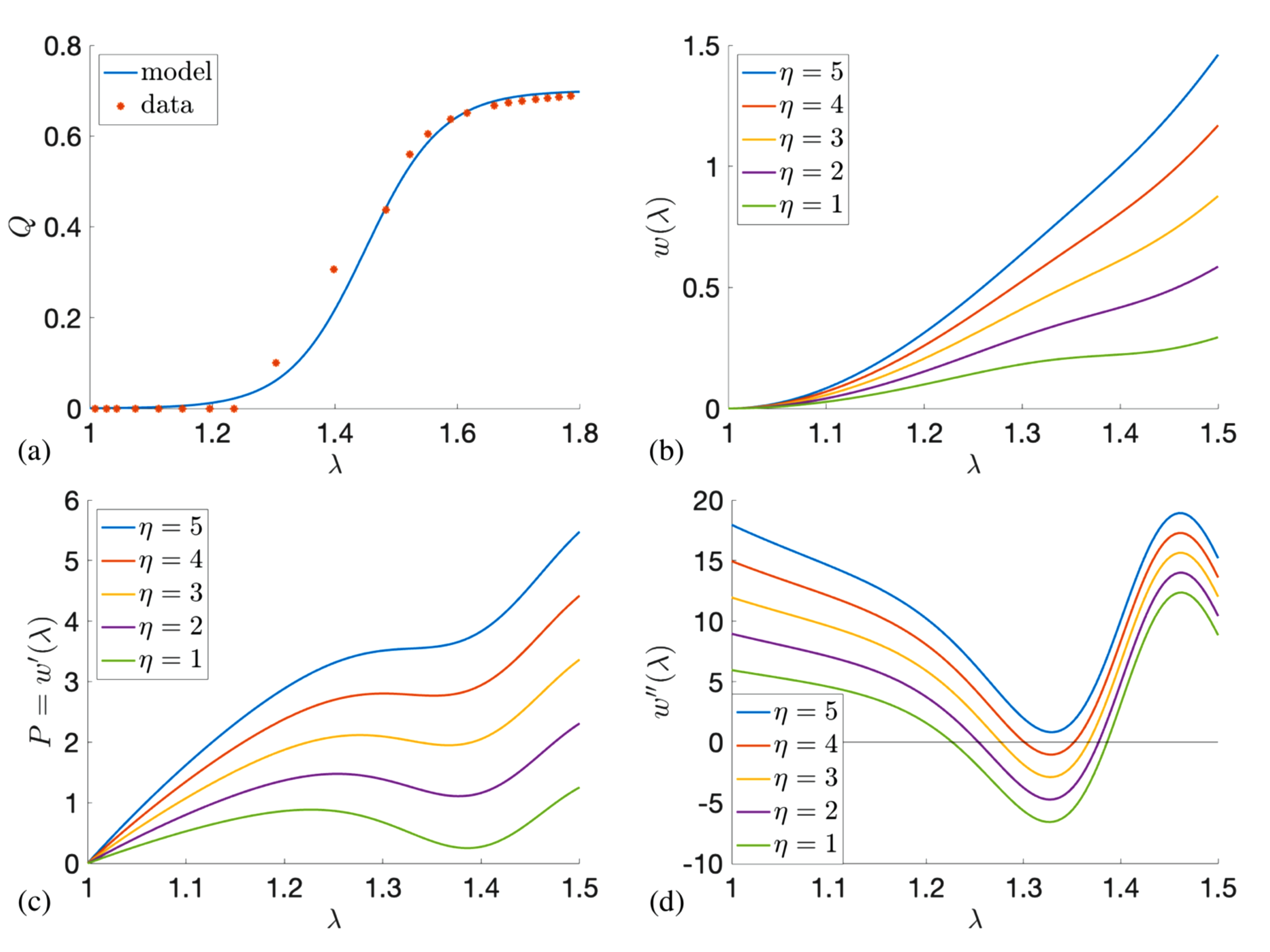}
		\caption{(a) The uniaxial order parameter $Q$ given by equation \eqref{eq:necking:Q} and the associated experimental data \cite{Fridrikh:1999:FT}; (b) The nondimensionalised strain-energy function $w(\lambda)$, for different values of $\eta=\mu^{(1)}/\mu^{(2)}$; (c) The applied tensile stress $P=w'(\lambda)$; (d) The second derivative $w''(\lambda)$ of the strain-energy function $w(\lambda)$.}\label{fig:necking}
	\end{center}
\end{figure}
%%%%%%%%%%%%%%%

We denote by $w(\lambda)$ the strain-energy function given by equation \eqref{eq:Wlce:nh:nc} with $\lambda$ as the only variable. Then the first derivative of this function is equal to the applied tensile stress, i.e.,
\begin{equation}
	w'(\lambda)=\left(P_{1}+p\lambda_{1}^{-1}\right)\frac{\partial\lambda_{1}}{\partial\lambda}+p\lambda_{2}^{-1}\frac{\partial\lambda_{2}}{\partial\lambda}+p\lambda_{3}^{-1}\frac{\partial\lambda_{3}}{\partial\lambda}=P,
\end{equation}
and necking forms when the applied stress $P$ first increases, then decreases, then increases again as $\lambda$ increases \cite{Mihai:2022}. It can be shown that, when $Q$ is constant, $w''(\lambda)>0$, therefore necking cannot occur if $Q$ is independent of deformation. A more general result is proved in Appendix~\ref{append:necking} of this paper. 

We let $Q$ depend on the deformation through the following relation obtained by calibration to experimental data of \cite{Fridrikh:1999:FT}, as shown in Figure~\ref{fig:necking}(a) (see also Appendix~\ref{append:Qdata} of this paper):
\begin{equation}\label{eq:necking:Q}
	Q=0.35\left[\frac{e^{16(\lambda-1.45)}-1}{e^{16(\lambda-1.45)}+1}-\frac{e^{16}-1}{e^{16}+1}+2\right],\qquad\mbox{for} \ \lambda\geq1.
\end{equation} 

It remains to verify numerically that the second derivative $w''(\lambda)$ of $w(\lambda)$ will change sign for some $\lambda\in(1,2)$. For different values of $\eta$, the nondimensionalised function $w(\lambda)$ (i.e., the function divided by $\mu^{(2)}$) and its first two derivatives are plotted in Figure~\ref{fig:necking}(b)-(d), respectively. These plots suggest that, as $\eta$ increases, the necking interval (i.e., the interval where $w''(\lambda)<0$) decreases, and there is a critical value $\eta_{crt}\approx4.55$ above which necking does not form. For example, when $\eta=\mu^{(1)}/\mu^{(2)}=3$, $w''(\lambda)$ changes from positive to negative at $\lambda_{*}\approx1.28$, then from negative to positive at $\lambda^{*}\approx1.37$. In this case, necking occurs for $1.28<\lambda<0.37$. When $\eta\to\infty$, the model reduces to the neo-Hookean model for rubber, which does not exhibit necking.

%%%%%%%%%%%%%%%%%%%%%%%%%%%%%%%%%%%%%%%%%%%%%%%%%%%%%%%%%%%%
%%%%%%%%%%%%%%%%%%%%   NEW SECTION  %%%%%%%%%%%%%%%%%%%%%%%%
%%%%%%%%%%%%%%%%%%%%%%%%%%%%%%%%%%%%%%%%%%%%%%%%%%%%%%%%%%%%
\section{Shear striping}\label{sec:stripes}

Next, we analyse shear striping under uniaxial stress, where $\lambda_{1}=\lambda$ and $\lambda_{2}=\lambda_{3}=\lambda^{-1/2}$. In this case, we assume that the nematic director only rotates in the biaxial plane while the uniaxial order parameter $Q$ remains constant. This assumption about the order parameter is consistent with experimental observations showing relatively small variations before, during and after stripe formation \cite{Higaki:2013:HTU,Kundler:1995:KF,Talroze:1999:etal,Zubarev:1999:etal}. 

In a Cartesian system of coordinates, we define the nematic director in the reference and stretched configuration, respectively, as 
\begin{equation}
	\textbf{n}_{0}=
	\left[
	\begin{array}{c}
		1\\
		0\\
		0
	\end{array}
	\right],\qquad
	\textbf{n}=
	\left[
	\begin{array}{c}
		\cos\theta\\
		\sin\theta\\
		0
	\end{array}
	\right],
\end{equation}
where $\theta\in[0,\pi/2]$ is the angle between $\textbf{n}$ and $\textbf{n}_{0}$. The associated natural deformation tensors take the form, respectively,
\begin{equation}
	\textbf{G}_{0}=
	\left[
	\begin{array}{ccc}
		a^{1/3} &  0 & 0 \\
		0 & a^{-1/6} & 0 \\
		0 & 0 & a^{-1/6}
	\end{array}
	\right]
\end{equation}
and
\begin{equation}
	\textbf{G}=
	\left[
	\begin{array}{ccc}
		a^{-1/6} +\left(a^{1/3}-a^{-1/6}\right)\cos^{2}\theta &  \left(a^{1/3}-a^{-1/6}\right)\sin\theta\cos\theta & 0 \\
		\left(a^{1/3}-a^{-1/6}\right)\sin\theta\cos\theta  & a^{-1/6} +\left(a^{1/3}-a^{-1/6}\right)\sin^{2}\theta & 0 \\
		0 & 0 & a^{-1/6}
	\end{array}
	\right].
\end{equation}
To demonstrate shear striping, following the procedure in \cite{Mihai:2023,Mihai:2020a:MG,Mihai:2021a:MG}, we consider the following perturbed deformation gradient 
\begin{equation}\label{eq:F:stripes:uniaxial}
	\textbf{F}=\left[
	\begin{array}{ccc}
		\lambda^{-1/2} &  0 & 0 \\
		\varepsilon & \lambda & 0 \\
		0 & 0 & \lambda^{-1/2}
	\end{array}
	\right],
\end{equation}
where $\lambda>1$ is the stretch ratio in the applied tensile force direction, and $0<\varepsilon\ll 1$ is a small shear parameter. The elastic deformation tensor is then equal to
\begin{equation}
	\begin{split}
		\textbf{A}&=\left[
		\begin{array}{ccc}
			\lambda^{-1/2}\left(a^{1/2}\sin^{2}\theta+\cos^{2}\theta\right)  & \lambda\left(a^{-1/2}-1\right)\sin\theta\cos\theta & 0 \\
			\lambda^{-1/2}\left(1-a^{1/2}\right)\sin\theta\cos\theta & \lambda\left(a^{-1/2}\sin^{2}\theta+\cos^{2}\theta\right) & 0 \\
			0 & 0 & \lambda^{-1/2}
		\end{array}
		\right]\\
		&+\varepsilon\left[
		\begin{array}{ccc}
			\left(1-a^{1/2}\right)\sin\theta\cos\theta & 0 & 0 \\
			\left(\sin^{2}\theta+a^{1/2}\cos^{2}\theta\right) & 0 & 0\\
			0 & 0 &0
		\end{array}
		\right].
	\end{split}
\end{equation}

We denote by $w(\lambda,\varepsilon,\theta)$ the strain-energy function described by equation \eqref{eq:Wlce:nh:nc}, depending only on $\lambda$, $\varepsilon$ and $\theta$. This function takes the following form,
\begin{equation}\label{eq:stripes:W}
	\begin{split}
		w(\lambda,\varepsilon,\theta)&=\frac{\mu^{(1)}}{2}\left(\lambda^2+2\lambda^{-1}+\varepsilon^2-3\right)\\
		&+\frac{\mu^{(2)}}{2}\left\{\left[\lambda^{-1/2}\left(a^{1/2}\sin^{2}\theta+\cos^{2}\theta\right)+\varepsilon\left(1-a^{1/2}\right)\sin\theta\cos\theta\right]^2\right.\\ 
		&+\left.\left[\lambda^{-1/2}\left(1-a^{1/2}\right)\sin\theta\cos\theta+\varepsilon\left(\sin^{2}\theta+a^{1/2}\cos^{2}\theta\right)\right]^2\right.\\
		&+\left.\left[\lambda\left(a^{-1/2}-1\right)\sin\theta\cos\theta\right]^2+\left[\lambda\left(a^{-1/2}\sin^{2}\theta+\cos^{2}\theta\right) \right]^2+\lambda^{-1}-3\right\}.
	\end{split}
\end{equation}
Differentiating with respect to $\varepsilon$ and $\theta$, respectively, gives
\begin{equation}\label{eq:dW:deps}
	\begin{split}
		\frac{\partial w(\lambda,\varepsilon,\theta)}{\partial\varepsilon}
		&=\mu^{(1)}\varepsilon\\
		&+\mu^{(2)}\left\{\left[\lambda^{-1/2}\left(a^{1/2}\sin^{2}\theta+\cos^{2}\theta\right)+\varepsilon\left(1-a^{1/2}\right)\sin\theta\cos\theta\right]\left(1-a^{1/2}\right)\sin\theta\cos\theta\right.\\
		&\left.+\left[\lambda^{-1/2}\left(1-a^{1/2}\right)\sin\theta\cos\theta+\varepsilon\left(\sin^{2}\theta+a^{1/2}\cos^{2}\theta\right) \right]\left(\sin^{2}\theta+a^{1/2}\cos^{2}\theta\right)\right\}
	\end{split}
\end{equation}
and
\begin{equation}\label{eq:dW:dtheta}
	\begin{split}
		\frac{\partial w(\lambda,\varepsilon,\theta)}{\partial\theta}&=\mu^{(2)}\left\{\left(a^{1/2}-1\right)\left[2\lambda^{-1/2}\sin\theta\cos\theta+\varepsilon\left(\sin^2\theta-\cos^{2}\theta\right)\right]\right.\\
		&\left.\cdot\left[\lambda^{-1/2}\left(a^{1/2}\sin^{2}\theta+\cos^{2}\theta\right)+\varepsilon\left(1-a^{1/2}\right)\sin\theta\cos\theta\right]\right.\\
		&\left.+\left(a^{1/2}-1\right)\left[\lambda^{-1/2}\left(\sin^{2}\theta-\cos^{2}\theta\right)-2\varepsilon\sin\theta\cos\theta\right]\right.\\
		&\left.\cdot\left[\lambda^{-1/2}\left(1-a^{1/2}\right)\sin\theta\cos\theta+\varepsilon\left(\sin^{2}\theta+a^{1/2}\cos^{2}\theta\right)\right]\right.\\
		&\left.+\lambda^2\left(a^{-1}-1\right) \sin\theta\cos\theta\right\}.
	\end{split}
\end{equation}
The equilibrium solution satisfies the simultaneous equations for energy minimisation,
\begin{equation}\label{eq:dwdepstheta}
	\frac{\partial w(\lambda,\varepsilon,\theta)}{\partial\varepsilon}=0\qquad\mbox{and}\qquad
	\frac{\partial w(\lambda,\varepsilon,\theta)}{\partial\theta}=0.
\end{equation}

At $\varepsilon=0$ and $\theta=0$, the partial derivatives defined by equations \eqref{eq:dW:deps}-\eqref{eq:dW:dtheta} are equal to zero, i.e., this solution is always an equilibrium state. For sufficiently small values of $\varepsilon$ and $\theta$, we can write the second order approximation 
\begin{equation}\label{eq:W:2order}
	w(\lambda,\varepsilon,\theta)\approx w(\lambda,0,0)+\frac{1}{2}\left(\varepsilon^2\frac{\partial^2 w}{\partial\varepsilon^2}(\lambda,0,0)+2\varepsilon\theta\frac{\partial^2 w}{\partial\varepsilon\partial\theta}(\lambda,0,0)+\theta^2\frac{\partial^2 w}{\partial\theta^2}(\lambda,0,0)\right),
\end{equation}
where
\begin{eqnarray}
	&&\frac{\partial^2 w}{\partial\varepsilon^2}(\lambda,0,0)=\mu^{(1)}+\mu^{(2)}a,\\
	&&\frac{\partial^2 w}{\partial\varepsilon\partial\theta}(\lambda,0,0)=\mu^{(2)}\lambda^{-1/2}\left(1-a\right),\\
	&&\frac{\partial^2 w}{\partial\theta^2}(\lambda,0,0)=\mu^{(2)}\left(\lambda^2-\lambda^{-1}a\right)\left(a^{-1}-1\right).
\end{eqnarray}
First, we find the equilibrium value $\theta_{0}$ for $\theta$ as a function of $\varepsilon$ by solving the second equation in \eqref{eq:dwdepstheta}. By the approximation \eqref{eq:W:2order}, the corresponding equation takes the form
\begin{equation}
	\varepsilon\frac{\partial^2 w}{\partial\varepsilon\partial\theta}(\lambda,0,0)+\theta\frac{\partial^2 w}{\partial\theta^2}(\lambda,0,0)=0,
\end{equation}
implying
\begin{equation}
	\theta_{0}(\varepsilon)=-\varepsilon\frac{\partial^2 w}{\partial\varepsilon\partial\theta}(\lambda,0,0)/\frac{\partial^2 w}{\partial\theta^2}(\lambda,0,0).
\end{equation}
Next, substituting $\theta=\theta_{0}(\varepsilon)$ in \eqref{eq:W:2order} yields the following function in $\varepsilon$,
\begin{equation}\label{eq:stability}
	w(\lambda,\varepsilon,\theta_{0}(\varepsilon))-w(\lambda,0,0)\approx\frac{\varepsilon^2}{2}\left[\frac{\partial^2 w}{\partial\varepsilon^2}(\lambda,0,0)-\left(\frac{\partial^2 w}{\partial\varepsilon\partial\theta}(\lambda,0,0)\right)^2/\frac{\partial^2 w}{\partial\theta^2}(\lambda,0,0)\right].
\end{equation}
Depending on whether the expression on the right-hand side is negative, zero, or positive, the respective equilibrium state is unstable, neutrally stable, or stable. We deduce that the equilibrium state with $\varepsilon=0$ and $\theta=0$ is unstable if 
\begin{equation}\label{eq:lambda:bound1:cp}
	a^{1/3}\left(\frac{\eta+1}{\eta+a}\right)^{1/3}<\lambda<a^{1/3},
\end{equation}
where $\eta=\mu^{(1)}/\mu^{(2)}$.

Similarly, at $\varepsilon=0$ and $\theta=\pi/2$, both the partial derivatives defined by \eqref{eq:dW:deps}-\eqref{eq:dW:dtheta} are equal to zero, and
\begin{eqnarray}
	&&\frac{\partial^2 w}{\partial\varepsilon^2}(\lambda,0,\pi/2)=\mu^{(1)}+\mu^{(2)},\\
	&&\frac{\partial^2 w}{\partial\varepsilon\partial\theta}(\lambda,0,\pi/2)=\mu^{(2)}\lambda^{-1/2}\left(a-1\right),\\
	&&\frac{\partial^2 w}{\partial\theta^2}(\lambda,0,\pi/2)=\mu^{(2)}\left(\lambda^{2}-\lambda^{-1}a\right)\left(1-a^{-1}\right).
\end{eqnarray}
Thus the equilibrium state with $\varepsilon=0$ and $\theta=\pi/2$ is unstable if 
\begin{equation}\label{eq:lambda:bound2:cp}
	a^{1/3}<\lambda<a^{1/3}\left(\frac{\eta+a}{\eta+1}\right)^{1/3}.
\end{equation}
Therefore, shear stripes can form when $\lambda$ satisfies  (see also partial results in the Appendix of \cite{Mihai:2023})
\begin{equation}\label{eq:stripes:bounds}
	a^{1/3}\left(\frac{\eta+1}{\eta+a}\right)^{1/3}<\lambda<a^{1/3}\left(\frac{\eta+a}{\eta+1}\right)^{1/3}.
\end{equation}

For  the corresponding equilibrium solution, the shear parameter and director angle take the following form, respectively:
\begin{eqnarray}
	&&\varepsilon_{0}=\pm\frac{\sqrt{\left[\lambda^3\left(\eta+a\right)-a\left(\eta+1\right)\right]\left[a\left(\eta+a\right)-\lambda^3\left(\eta+1\right)\right]}}{a\lambda^{1/2}\left(2\eta+a+1\right)},\label{eq:eps0}\\
	&&\theta_{0}=\pm\arctan\sqrt{\frac{\lambda^3\left(\eta+a\right)-a\left(\eta+1\right)}{a\left(\eta+a\right)-\lambda^3\left(\eta+1\right)}}.\label{eq:theta0}
\end{eqnarray} 

The gradient tensors for alternating shear deformations in two adjacent stripe subdomains are $\textbf{F}_{\pm}$ with $\varepsilon=\pm\varepsilon_{0}$, respectively. These two deformations are geometrically compatible in the sense that $\textbf{F}_{+}$ and $\textbf{F}_{-}$ are rank-one connected, i.e., $\mathrm{rank}\left(\textbf{F}_{+}-\textbf{F}_{-}\right)=1$ \cite{Mihai:2023:MG}. 

%%%%%%%%%%%%%%%%
\begin{figure}[htbp]
	\begin{center}
		\includegraphics[width=0.5\textwidth]{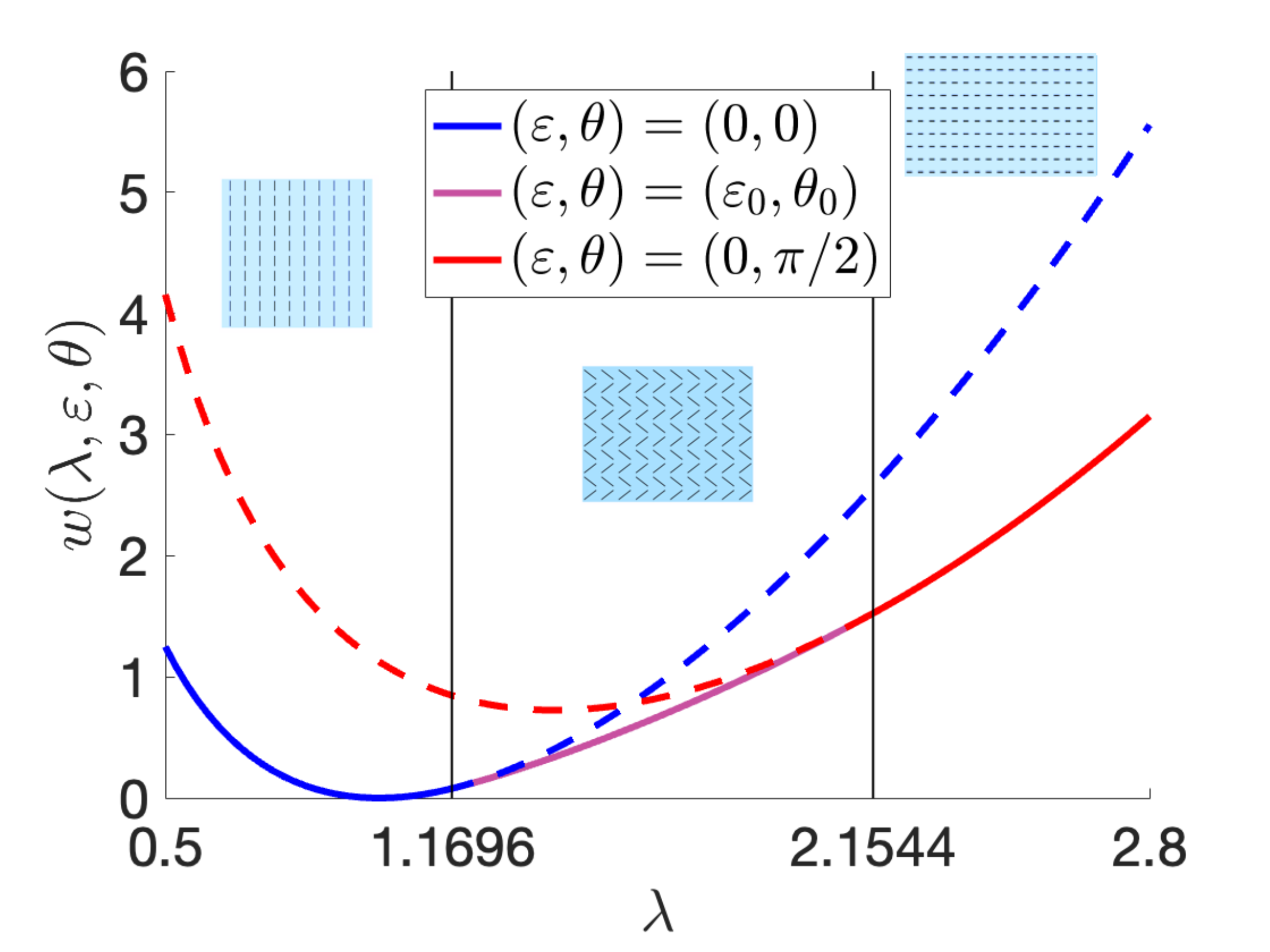}
		\caption{The nondimensionalised strain-energy function $w(\lambda,\varepsilon,\theta)$ given by equation \eqref{eq:stripes:W}, for $(\varepsilon,\theta)=(0,0)$, $(\varepsilon,\theta)=(\varepsilon_{0},\theta_{0})$, and $(\varepsilon,\theta)=(0,\pi/2)$, when $a=4$ and $\eta=1$ (i.e., $\mu^{(1)}=\mu^{(2)}$). The two vertical lines correspond to the lower and upper bounds on $\lambda$. Between these bounds, the second solution, with $(\varepsilon,\theta)=(\varepsilon_{0},\theta_{0})$, minimises the energy.}\label{fig:stripes-energy}
	\end{center}
\end{figure}
%%%%%%%%%%%%%%%%

%%%%%%%%%%%%%%%%%
\begin{figure}[htbp]
	\begin{center}
		\includegraphics[width=0.99\textwidth]{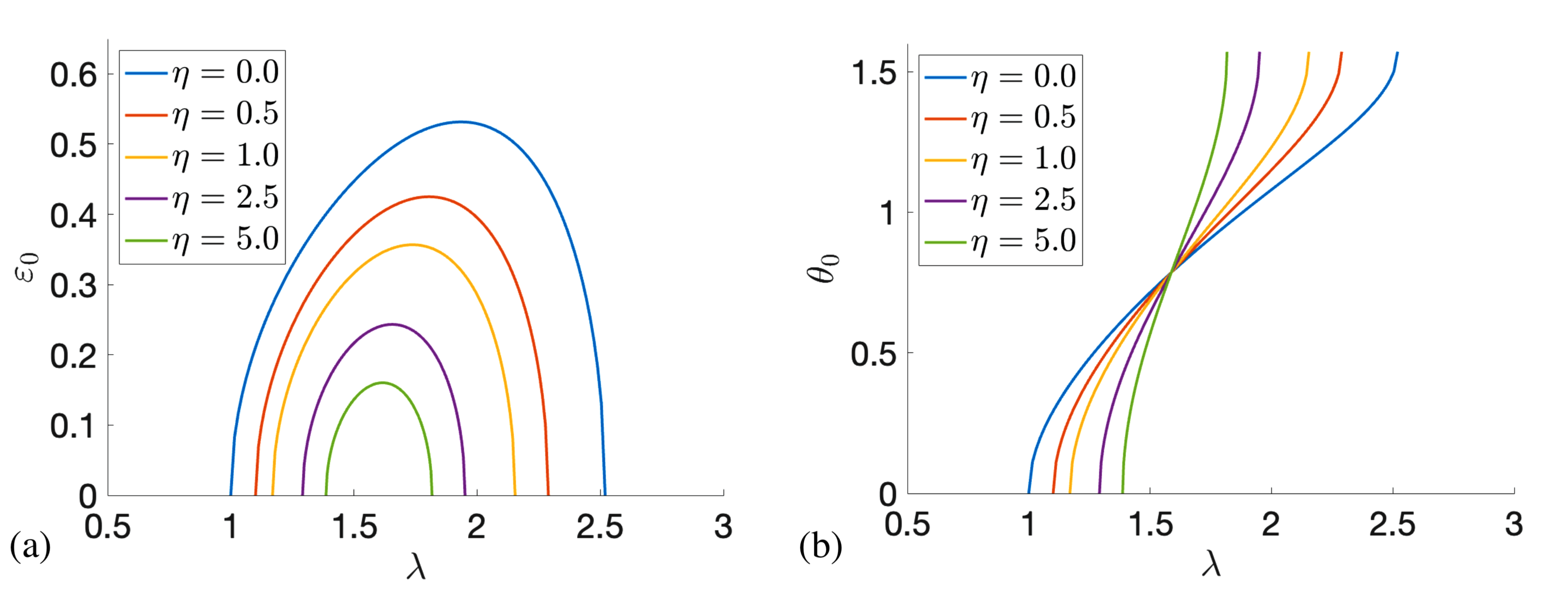}
		\caption{The positive values of the (a) shear parameter $\varepsilon_{0}$ and (b) director angle $\theta_{0}$ given by equations \eqref{eq:eps0}-\eqref{eq:theta0}, for different values of the parameter ratio $\eta=\mu^{(1)}/\mu^{(2)}$ when $a=4$. }\label{fig:eps0theta0}
	\end{center}
\end{figure}
%%%%%%%%%%%%%%%

The nondimensionalised strain-energy function $w(\lambda,\varepsilon,\theta)$ (i.e., the function divided by $\mu^{(2)}$) described by equation \eqref{eq:stripes:W}, with $\eta=\mu^{(1)}/\mu^{(2)}=1$, is illustrated in Figure~\ref{fig:stripes-energy}. For $\lambda$ with values between the lower and upper bounds given by \eqref{eq:stripes:bounds}, the minimum energy is attained for $(\varepsilon,\theta)=(\varepsilon_{0},\theta_{0})$ provided by equations \eqref{eq:eps0}-\eqref{eq:theta0}. Figure~\ref{fig:eps0theta0} depicts the positive values of the shear parameter $\varepsilon_{0}$ and director angle $\theta_{0}$ described by equations \eqref{eq:eps0} and \eqref{eq:theta0}, respectively, for different parameter ratios $\eta$. When $\eta=0$, corresponding to the neoclassical form, shear stripes form for $\lambda\in\left(1,a^{2/3}\right)$. When $\eta\to\infty$, there is no shear striping since the material is practically purely elastic. 

%%%%%%%%%%%%%%%%%%%%%%%%%%%%%%%%%%%%%%%%%%%%%%%%%%%%%%%%%%%%
\subsection{Optimal stripe width} 

For shear striping, the width of each subdomain depends on the material properties, sample geometry and boundary conditions. Dirichlet conditions fixing the director at both ends of the LCE sample in its initial orientation, i.e., perpendicular to the longitudinal direction, known as \emph{strong anchoring}, play an important role in numerically capturing this property. To estimate the stripe width, we add the Frank energy density to our model function, as follows,
\begin{equation}\label{eq:Wnlce:nh:nc:f}
	W^{(nlce)}=W^{(lce)}+\frac{\texttt{k}}{2}\left|\textbf{F}^{T}\mathrm{Grad}\ \textbf{n}\right|^2,
\end{equation}
where $W^{(lce)}$ is as defined in equation \eqref{eq:Wlce:nh:nc} and $\texttt{k}$ is the Frank constant \cite{deGennes:1993:dGP,Warner:2007:WT}.

We denote by $L_{1},  L_{2}, L_{3}$ the size of the undeformed LCE sample in the extension direction (which is perpendicular to the director orientation in the undeformed state), the width direction (which is parallel to the initial director), and the thickness direction, respectively. Provided that Dirichlet boundary conditions are imposed at the ends, the width $h$ of a strip band forming parallel to the initial director at each of these ends must be of the same order of magnitude as that of a shear stripe \cite{Verwey:1996:VWT} (see also \cite[Appendix~C]{Warner:2007:WT}).  For geometric compatibility, the displacement in such a strip band must be equal to $\gamma h$, where $\gamma=\lambda_{12}(x_{1},x_{2})$ denotes the shear strain in a shear stripe, so that
\begin{equation}
\frac{\mathrm{d}\lambda_{12}}{\mathrm{d}x_{2}}=\frac{\mathrm{d}\lambda_{22}}{\mathrm{d}x_{1}},
\end{equation}
with $\lambda_{22}\approx h$ representing the stretch ratio in the direction of the stripe width.

Due to the infinitesimal scale of the strip band width, the linear elastic framework is suitable to analyse their deformation. The corresponding elastic energy is then approximately
\begin{equation}
E_{end}\approx\frac{1}{2}\mu\gamma^2\left(2L_{3}h^2\right)\frac{L_{2}}{h}=\mu\gamma^2L_{2}L_{3}h,
\end{equation}
where $L_{3}h^2$ is the volume of one end region for a single shear stripe and $L_{2}/h$ is the number of  stripes. This elastic energy decreases as $h$ decreases.

There is additional elastic energy at the narrow interface between adjacent stripes with opposite shear deformation and director rotation. Since the director angle $\theta\in(-\theta_{0},\theta_{0})$ varies from one side of the interface to the other, so that it takes the value zero in the middle of the interval, this energy is approximately
\begin{equation}
E_{int}\approx L_{1}L_{3}\sqrt{\mu\texttt{k}}\left(\frac{L_{2}}{h}-1\right),
\end{equation}
where $\sqrt{\mu\texttt{k}}$ is the effective surface tension, $L_{1}L_{3}$ is the area of a single interface, and $L_{2}/h-1$ is the number of interfaces. This energy tends to decrease when $h$ increases.

The problem is then to find the optimal stripe width $h$ that minimises the total energy
\begin{equation}
E_{stripe}=E_{end}+E_{int}.
\end{equation}
Differentiating the above function with respect to $h$ and solving for the critical value, we obtain 
\begin{equation}\label{eq:hopt}
h\approx \left(L\sqrt{\frac{\texttt{k}}{\mu}}\right)^{1/2},
\end{equation}
i.e., the stripe width increases with sample initial length $L_{1}=L$ and nematic penetration depth $\sqrt{\texttt{k}/\mu}\approx 10^{-8}$ m, which measures the competition between polymer and Frank elasticity. A typical value for the stripe width is $h\approx 10^{-5}$ m. 

%%%%%%%%%%%%%%%%%%%%%%%%%%%%%%%%%%%%%%%%%%%%%%%%%%%%%%%%%%%%
%%%%%%%%%%%%%%%%%%%%   NEW SECTION  %%%%%%%%%%%%%%%%%%%%%%%%
%%%%%%%%%%%%%%%%%%%%%%%%%%%%%%%%%%%%%%%%%%%%%%%%%%%%%%%%%%%%
\section{Numerical simulations}\label{sec:numerics}

We illustrate numerically first the stretch deformation causing necking, then the periodic shear striping for nematic LCEs. 

%%%%%%%%%%%%%%%%
\begin{figure}[htbp]
	\begin{center}
		\includegraphics[width=0.9\textwidth]{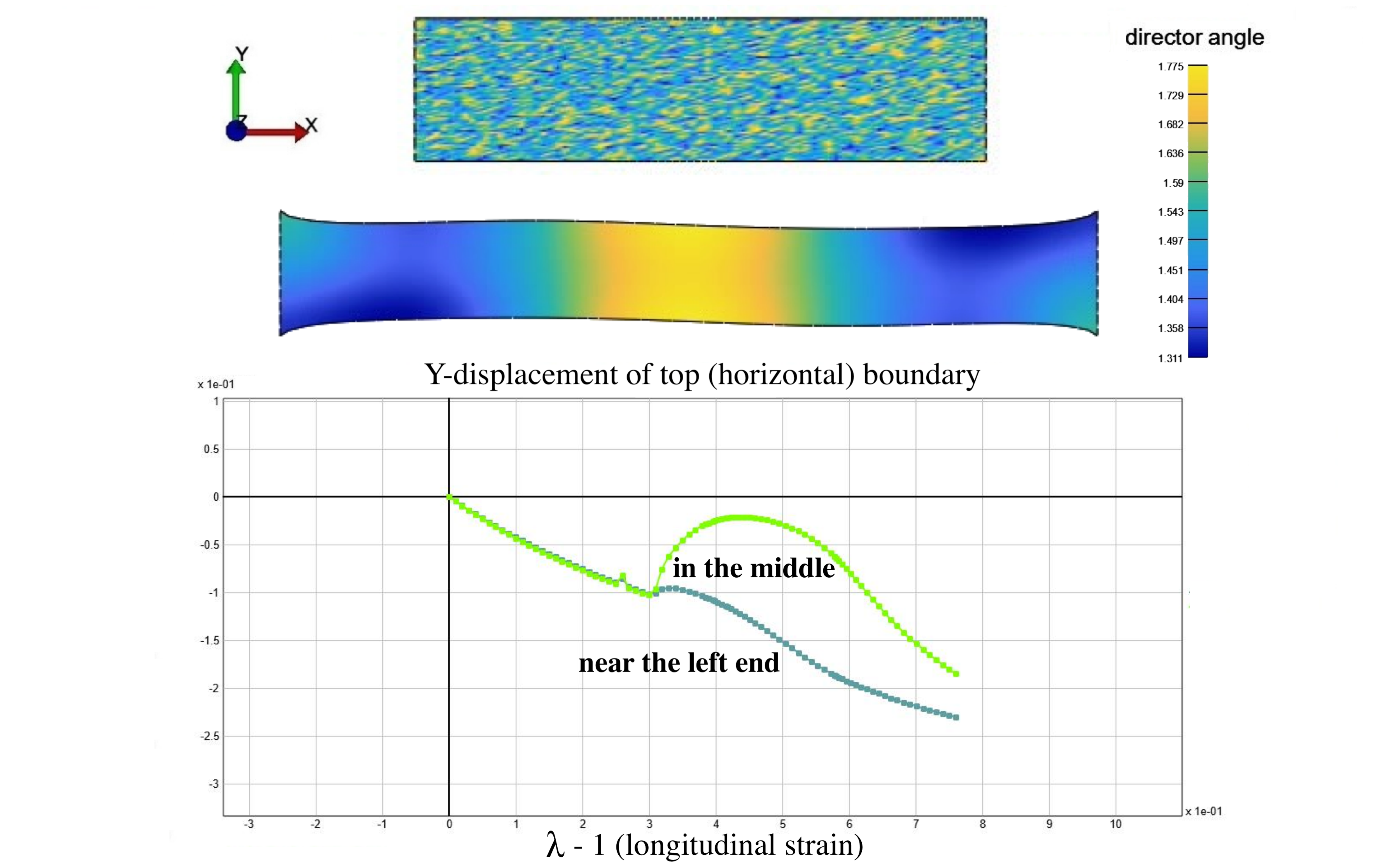}
		\caption{Finite element simulation of necking in a nematic LCE described by equation \eqref{eq:Wlce:nh:nc} when $\eta=\mu^{(1)}/\mu^{(2)}=3$. Top image shows the reference configuration, middle image shows the deformed state with the director orientation, and bottom plot is for the vertical displacement in the middle and near the left end of the top horizontal boundary.}\label{fig:nematic-necking-febio}
	\end{center}
\end{figure}
%%%%%%%%%%%%%%%

%%%%%%%%%%%%%%%%
\begin{figure}[htbp]
	\begin{center}
		\includegraphics[width=0.85\textwidth]{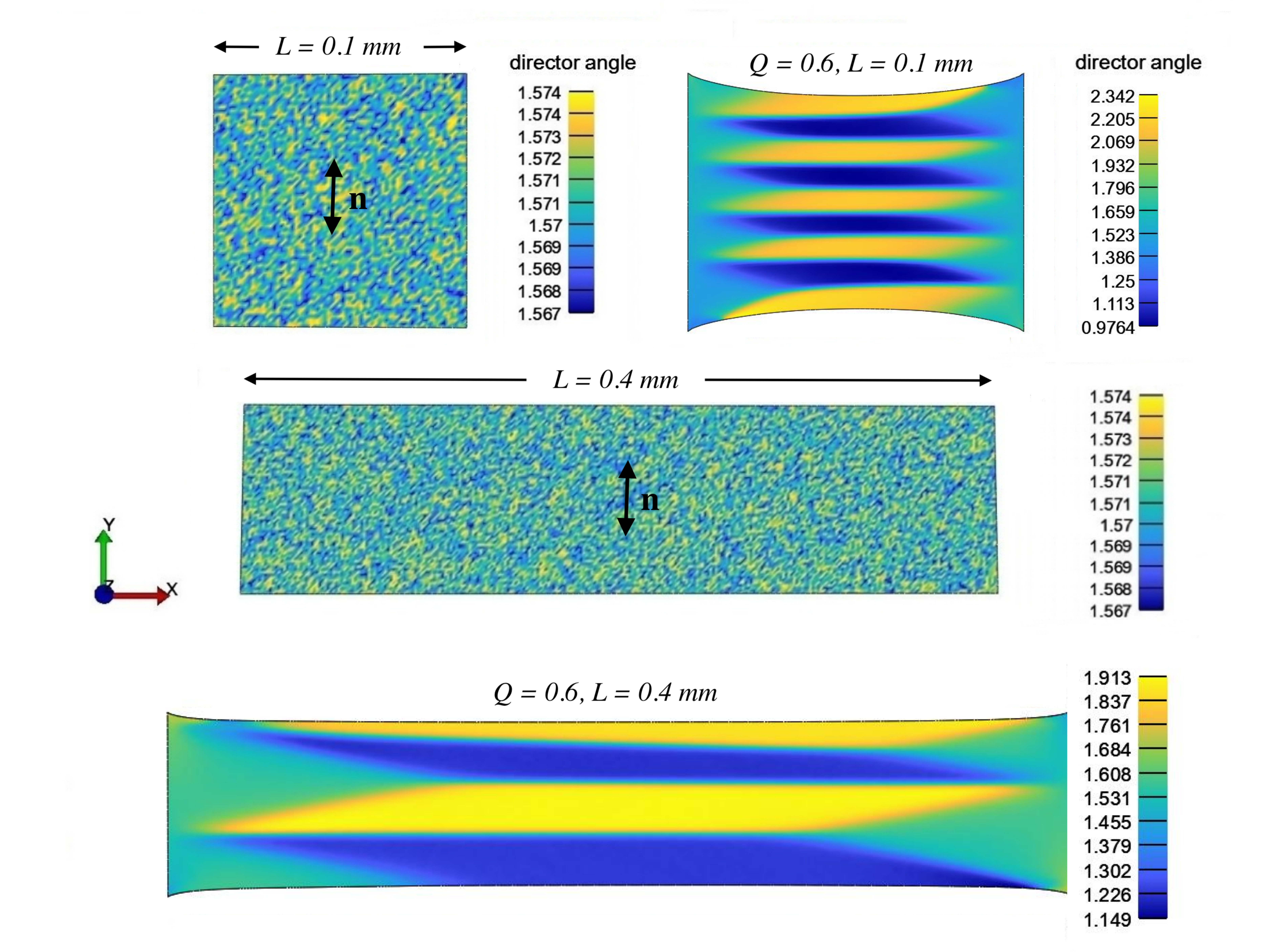}
		\caption{Finite element simulation of nematic LCEs described by equation \eqref{eq:Wnlce:nh:nc:f}, with $\eta=\mu^{(1)}/\mu^{(2)}=1$, where stripe width increases with the sample initial length $L$. Both the reference and deformed configurations are shown.}\label{fig:nematic-stripeslength-febio}
	\end{center}
\end{figure}
%%%%%%%%%%%%%%%

%%%%%%%%%%%%%%%%
\begin{figure}[htbp]
	\begin{center}
		\includegraphics[width=0.85\textwidth]{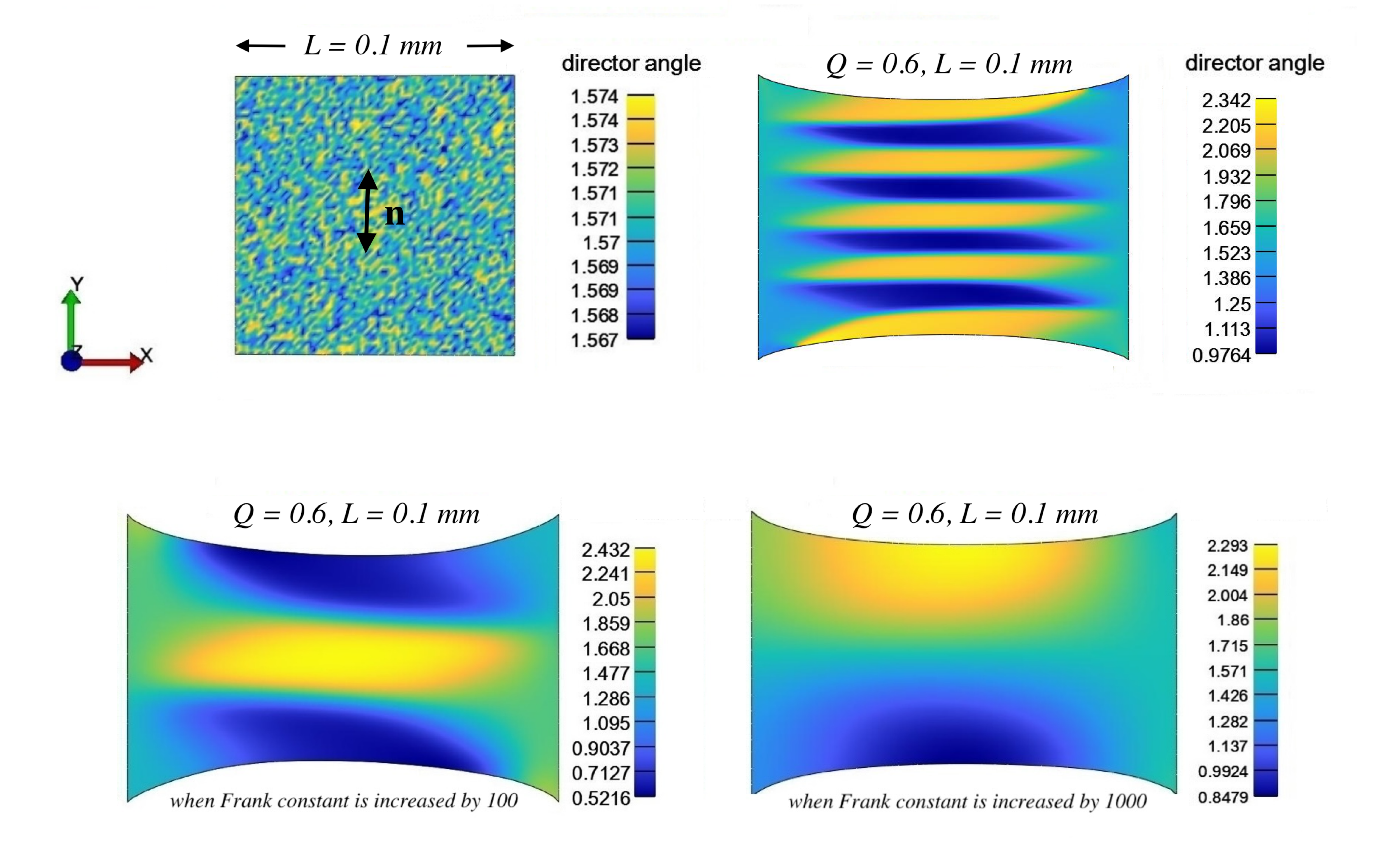}
		\caption{Finite element simulation of nematic LCEs described by equation \eqref{eq:Wnlce:nh:nc:f}, with $\eta=\mu^{(1)}/\mu^{(2)}=1$, where stripe width increases with the nematic penetration depth $\sqrt{\texttt{k}/\mu}$ ($\mu=\mu^{(1)}+\mu^{(2)}$). Both the reference and deformed configurations are shown.}\label{fig:nematic-stripesfrank-febio}
	\end{center}
\end{figure}
%%%%%%%%%%%%%%%

Figure~\ref{fig:nematic-necking-febio} presents a finite element simulation of necking where the model function is given by equation  \eqref{eq:Wlce:nh:nc} with $3\mu^{(1)}= \mu^{(2)}=10^{4}$ Pa. In this figure, the deformed configuration appears narrower near the ends than in the middle. This is confirmed by the plots showing the corresponding vertical displacements as stretching increases. The darker colour at the ends of the deformed state further suggest that the director is less aligned with the tensile direction compared to the middle section.

Figures~\ref{fig:nematic-stripeslength-febio}-\ref{fig:nematic-stripesfrank-febio} show finite element examples of shear striping in a nematic LCE described by equation \eqref{eq:Wnlce:nh:nc:f} with $\mu^{(1)}= \mu^{(2)}=5\cdot 10^{4}$ Pa and $\texttt{k}=10^{-11}$ N. In Figure~\ref{fig:nematic-stripeslength-febio}, for two samples with different aspect ratios, the number of stripes reduces almost by half when the initial length $L$ increases 4 times. In Figure~\ref{fig:nematic-stripesfrank-febio}, the number of stripes reduces by $\approx3$ when Frank constant $\texttt{k}$ increases 100 times, and by $\approx5$ when Frank constant  increases 1,000 times. This is in agreement with the estimate of equation \eqref{eq:hopt}.

The computed examples presented here were realised by building on existing procedures for three-dimensional finite elasticity of the open-source software Finite Elements for Biomechanics (FEBio) \cite{Maas:2012:MEAW}. In addition, our numerical implementation accounts also for macroscopic nematic effects in continuum LCE models \cite{Luo:2012:LC}.  Under uniaxial stretch \cite{Warner:2007:WT}, Dirichlet boundary conditions are imposed on each side parallel to the $Y$-directions, where the displacement in all directions is prescribed and the nematic director is fixed, while all the other boundaries are free.

%%%%%%%%%%%%%%%%%%%%%%%%%%%%%%%%%%%%%%%%%%%%%%%%%%%%%%%%%%%%
%%%%%%%%%%%%%%%%%%%%   NEW SECTION  %%%%%%%%%%%%%%%%%%%%%%%%
%%%%%%%%%%%%%%%%%%%%%%%%%%%%%%%%%%%%%%%%%%%%%%%%%%%%%%%%%%%%
\section{Conclusion}

We have demonstrate theoretically and numerically deformation localisation through necking and shear striping in nematic elastomer models. These two inherent instability mechanisms  appear in certain nematics LCEs when subject to uniaxial tensile stress. For necking, the scalar uniaxial order parameter varies, reflecting the polydomain-monodomain transition under mechanical stretch. For shear striping, the order parameter is fixed.  Our finite element simulations capture well both of these phenomena. The modelling approaches adopted here can further be extended to include probabilistic model parameters \cite{Mihai:2022} and strain rate effects \cite{Chehade:2024:etal,Zhang:2019:ZXJH}.

\vskip16pt

%\noindent\textbf{Funding details:} We are grateful for the PhD funding support by Cardiff University to Rabin Poudel.\\

%\noindent\textbf{Data availability statement:} The data for this research are openly available at https://doi.org/10.26078/6170-d520\\

%\noindent\textbf{Disclosure statement:} The authors declare that they have no competing interests.\\

%%%%%%%%%%%%%%%%%%%%%%%%%%%%%%%%%%%%%%%%%%%%%%%%%%%%%%%%%%%%
%%%%%%%%%%%%%%%%%%%%   NEW SECTION  %%%%%%%%%%%%%%%%%%%%%%%%
%%%%%%%%%%%%%%%%%%%%%%%%%%%%%%%%%%%%%%%%%%%%%%%%%%%%%%%%%%%%
\appendix
\section{Nematic elastomer models where necking cannot form}\label{append:necking}
\setcounter{equation}{0}
\renewcommand{\theequation}{A.\arabic{equation}}

In this appendix, we consider the following strain-energy functions describing a nematic LCE,
\begin{equation}\label{eq:Wnc}
	W^{(nc)}(\textbf{F},\textbf{n})=W(\textbf{A}),
\end{equation}
where $\textbf{F}$ represents the macroscopic deformation gradient from the cross-linking state, $\textbf{n}$ is the nematic director in the present configuration, and $W(\textbf{A})$ is the strain-energy density of the isotropic polymer network, depending only on the (local) elastic deformation tensor $\textbf{A}=\textbf{G}^{-1}\textbf{F}\textbf{G}_{0}$, with $\textbf{G}_{0}$ and $\textbf{G}$ the natural deformation tensors due to the liquid crystal director in the reference and current configuration, respectively.

The LCE is subject to the finite-strain deformation with $\lambda_{1}=\lambda>1$ and $\lambda_{2}=\lambda_{3}=\lambda^{-1/2}$, while the tensile force is parallel to the nematic director. Let  $\textbf{G}_{0}=\textbf{I}$, so that 
\begin{equation}
	\alpha_{1}=\lambda\left(\frac{1+2Q}{1-Q}\right)^{-1/3}=\alpha,\qquad
	\alpha_{2}=\alpha_{3}=\lambda^{-1/2}\left(\frac{1-Q}{1+2Q}\right)^{-1/6}=\alpha^{-1/2},
\end{equation}
where $Q\in(0,1)$ is the scalar uniaxial order parameter in the current configuration.

For hyperelastic materials where the force-extension curve does not have a maximum, the homogeneous deformation is the only absolute minimiser of the elastic energy, hence necking cannot occur  \cite{Sivaloganathan:2011:SS}. Here, we show that, if $Q$ is constant and the hyperelastic model described by $W(\textbf{A})$ does not exhibit necking, then the LCE model $W^{(nc)}(\textbf{F})$ does not present necking either.  

To prove this, we denote by $w(\lambda)$ the strain-energy function $W^{(nc)}(\textbf{F})$ given by equation \eqref{eq:Wnc}, depending only on $\lambda$, and by $\widetilde{w}(\alpha)$  the strain-energy function $W(\textbf{A})$ for the underlying polymeric network, depending only on $\alpha$. Calculating the first and second derivatives, we obtain
\begin{equation}
	\frac{\mathrm{d}w(\lambda)}{\mathrm{d}\lambda}=\frac{\mathrm{d}\widetilde{w}(\alpha)}{\mathrm{d}\alpha}\frac{\mathrm{d}\alpha}{\mathrm{d}\lambda}=\frac{\mathrm{d}\widetilde{w}(\alpha)}{\mathrm{d}\alpha}\left(\frac{1+2Q}{1-Q}\right)^{-1/3}
\end{equation}
and
\begin{equation}
	\frac{\mathrm{d^2}w(\lambda)}{\mathrm{d}\lambda^2}=\frac{\mathrm{d}^2\widetilde{w}(\alpha)}{\mathrm{d}\alpha^2}\frac{\mathrm{d}\alpha}{\mathrm{d}\lambda}\left(\frac{1+2Q}{1-Q}\right)^{-1/3}=	\frac{\mathrm{d}^2\widetilde{w}(\alpha)}{\mathrm{d}\alpha^2}\left(\frac{1+2Q}{1-Q}\right)^{-2/3}>0.
\end{equation}
The last inequality holds since, by assumption, there is no necking for the polymeric network, hence $\mathrm{d}^2\widetilde{w}/\mathrm{d}\alpha^2>0$. It follows that $\mathrm{d}^2w/\mathrm{d}\lambda^2>0$, i.e., for the LCE described by equation \eqref{eq:Wnc}, necking will not form. 

In particular, since incompressible neo-Hookean \cite{Treloar:1944} and Mooney-Rivlin \cite{Mooney:1940,Rivlin:1948:IV} hyperelastic models do not exhibit necking, this property is inherited by the associated LCE models of the form given by equation \eqref{eq:Wnc}  (see also \cite{Mihai:2021a:MG} for alternative modelling). 

%%%%%%%%%%%%%%%%%%%%%%%%%%%%%%%%%%%%%%%%%%%%%%%%%%%%%%%%%%%%
%%%%%%%%%%%%%%%%%%%%   NEW SECTION  %%%%%%%%%%%%%%%%%%%%%%%%
%%%%%%%%%%%%%%%%%%%%%%%%%%%%%%%%%%%%%%%%%%%%%%%%%%%%%%%%%%%%
\section{Experimental data}\label{append:Qdata}
\setcounter{equation}{0}
\renewcommand{\thetable}{B.\arabic{table}}

We present in this appendix the experimental data for scalar uniaxial order parameter used to calibrate the function given in equation \eqref{eq:necking:Q}. These data are plotted in Figure~\ref{fig:necking}(a). 

%%%%%%%%%%%%%%%%%%%
\begin{table}[htbp] 
	\centering
	\renewcommand{\arraystretch}{1.5}
	\caption{Experimental data  \cite{Fridrikh:1999:FT}.}\label{table:Qdata}
	\vspace*{0.15cm}
	\scalebox{0.9}{
		\begin{tabular}{c|c}
			\hline
			$\lambda$ & $Q$ \\			
			longitudinal stretch & uniaxial order parameter \\
			\hline
			1.7853  &  0.6881\\
			1.7658  &  0.6860\\
			1.7476  &  0.6838\\
			1.7281  & 0.6808\\
			1.7061  & 0.6772\\
			1.6840 &  0.6727\\
			1.6606 &  0.6669\\
			1.6164  &  0.6514\\
			1.5891  &  0.6365\\
			1.5526 &   0.6050\\
			1.5227 & 0.5595\\
			1.4836 &   0.4378\\
			1.3989 &  0.3059\\
			1.3037 & 0.1000\\
			1.2359 &    0\\
			1.1955  &     0\\
			1.1512   &  0\\
			1.1121   &    0\\
			1.0743 &     0\\
			1.0443 &    0\\
			1.0274 &      0\\
			1.0091 &      0\\
			\hline
		\end{tabular}
	}
\end{table}
%%%%%%%%%%%%%%%%%%

%%%%%%%%%%%%%%%%%%%%%%%%%%%%%%%%%%%%%%%%%%%%%%%%%%%%%%%%%%%%
%%%%%%%%%%%%%%%%%%%%  NEW SECTION   %%%%%%%%%%%%%%%%%%%%%%%%
%%%%%%%%%%%%%%%%%%%%%%%%%%%%%%%%%%%%%%%%%%%%%%%%%%%%%%%%%%%%

%%%%%%%%%%%%%%%%%%%
\end{document}